\documentclass[12pt]{article}

  \usepackage{graphicx}
  \usepackage{epsfig}
  \usepackage{amssymb}
  \usepackage{subfigure}
  \usepackage{axodraw}

\evensidemargin - 1.truecm
\def\({\left(}
\def\){\right)}

\newcommand{\ep}{\varepsilon}
\newcommand{\Li}[2]{{\mbox{Li}}_{#1}\left(#2\right)}

\newcommand{\Ls}[2]{{\mbox{Ls}}_{#1}\left(#2\right)}

\newcommand{\be}{\begin{equation}}
\newcommand{\ee}{\end{equation}}
\newcommand{\nn}{\nonumber}
\newcommand{\bea}{\begin{eqnarray}}
\newcommand{\eea}{\end{eqnarray}}
\newcommand{\bfig}{\begin{figure}}
\newcommand{\efig}{\end{figure}}
\newcommand{\bc}{\begin{center}}
\newcommand{\ec}{\end{center}}
\newcommand{\bd}{\begin{displaymath}}
\newcommand{\ed}{\end{displaymath}}

\begin{document}

\begin{titlepage}
\nopagebreak
{\flushright{
        \begin{minipage}{5cm}
        Freiburg-THEP 04/17\\
        {\tt hep-ph/0410216}\\
        \end{minipage}        }

}
\vspace*{-1.5cm}                        
\vskip 3.5cm
\begin{center}
\boldmath
{\Large \bf Three-Loop   
  Electroweak Correction to the Rho Parameter in the Large Higgs Mass Limit}
  \unboldmath
\vskip 1.2cm
{\large R.~Boughezal,\footnote{Email: 
{\tt Radja.Boughezal@physik.uni-freiburg.de}}
J.B.~Tausk\footnote{Email: 
{\tt Tausk@physik.uni-freiburg.de}}
and J.J.~van~der~Bij\footnote{Email: 
{\tt jochum@physik.uni-freiburg.de}}} \\[2mm] 

\vskip .7cm
{\it  Fakult\"at f\"ur Mathematik und Physik, 
Albert-Ludwigs-Universit\"at
Freiburg, \\ D-79104 Freiburg, Germany} 
\vskip .3cm
 
\end{center}
\vskip 1.5cm


\begin{abstract} 

We present an analytical calculation of the leading three-loop 
radiative correction to the $\rho$-parameter in the
Standard Model in the large Higgs mass limit.
This correction, of order $g^6 m_H^4/M_W^4$, is opposite in
sign to the leading two-loop correction of order $g^4 m_H^2/M_W^2$.
The two corrections cancel each other for a Higgs mass of approximately
$480~\mbox{GeV}$. The result shows that it is extremely unlikely
that a strongly interacting Higgs sector could fit the data
of electroweak precision measurements.

\vskip .7cm 
\flushright{
        \begin{minipage}{12.3cm}
        \end{minipage}        }
\end{abstract}
\vfill
\end{titlepage}    


\section{Introduction}
\label{sec:intro}
The Standard Model of electroweak physics is in good agreement 
with the data. The only particle of the Standard Model, that has not
been detected so far is the Higgs boson. Direct searches give
a lower bound of $m_H=114.4~$GeV~\cite{Barate:2003sz}.
The precision of present day 
experiments even makes it possible to put limits on the Higgs
mass through its influence in radiative corrections.
At the one-loop level, such radiative corrections grow logarithmically
with the Higgs boson mass~\cite{Veltman:1976rt}. Fits to the
data imply a light Higgs boson. However the situation is
not completely satisfactory,
as there is difference of about $2.9~\sigma$ between the
two most precise determinations of the effective electroweak
mixing angle,
$\sin^2 \theta^{\mbox{lept}}_{\mbox{eff}}$
i.e. the one based on the measurement of the $b$-quark forward-backward
asymmetry $A^{0,b}_{\mbox{FB}}$ at LEP on one hand, and the one based
on measurements of the leptonic asymmetry parameter $A_{\ell}$ at SLD
on the other~\cite{Grunewald:2003ij,LEP}. The value of
the Higgs mass preferred by the b-quark data is around
$0.5~$TeV, while the leptonic asymmetry data and the
$W$-boson mass point to a value which is slightly below
the lower bound from the direct
searches~\cite{Chanowitz:2003hx,Ferroglia:2004jx}.
With the recent measurements
of the W-mass and top mass from the Tevatron this is well within
statistics. Still there is the logical possibility that the Higgs
boson is very heavy ($\approx 1~$TeV) and strongly interacting.
Since the Higgs self-coupling grows like $m_H^2$ higher loop effects
can play a role and cancel against the one-loop leading $\log(m_H)$
effects. At the two-loop level such radiative corrections have been
calculated, also allowing for anomalous Higgs-boson
self-couplings~\cite{vanderBij:1983bw,vanderBij2}. Inclusion of
these two-loop corrections does not make it possible to fit the
data with a heavy Higgs boson.
However in a recent paper it was shown that the two-loop
correction is accidentally anomalously small and therefore important
effects might first appear only at the three-loop level~\cite{akhoury}. The
situation can only be clarified by an explicit three-loop
calculation. This three-loop calculation for one of the precision
variables, the so-called $\rho$-parameter, is the subject of this paper.

The electroweak $\rho$-parameter is a measure of the relative strengths
of neutral and charged-current interactions in four-fermion processes
at zero momentum transfer~\cite{rhodef}. After defining the
Fermi constant $G_F$ by means of the effective interaction
that describes muon decay,
\begin{equation}
\label{eq:LCC}
{\cal L}^{CC} = - \frac{G_F}{\sqrt{2}}
 \left[ \bar{\nu}_{\mu} \gamma^{\mu} \( 1 + \gamma_5 \) \mu \right]
 \left[ \bar{e} \gamma_{\mu} \( 1 + \gamma_5 \) \nu_e \right] \,
\end{equation}
one can define $\rho$ and the sine of the weak mixing angle $s_W = \sin
\theta_W$ as the parameters that appear in the effective interaction
that describes the scattering of neutrinos and anti-neutrinos by electrons,
\begin{equation}
\label{eq:LNC}
{\cal L}^{NC} = - \frac{\rho \, G_F}{2\sqrt{2}}
 \left[ \bar{\nu}_{\mu} \gamma^{\mu} \( 1 + \gamma_5 \) \nu_{\mu} \right]
 \left[ \bar{e} \gamma_{\mu} \( 1 - 4 s_W^2 + \gamma_5 \) e \right] \, .
\end{equation}
In the Standard Model, at tree level, it is related to the $W$ and $Z$ boson
masses by
\begin{equation}
\label{eq:rhotree}
\rho = \frac{M_W^2}{c_W^2\,M_Z^2} = 1 \, ,
\end{equation}
where $c_W = \cos \theta_W$.
This relation gets modified by radiative corrections
\begin{equation}
\rho = \frac{1}{1 - \Delta\rho}
\end{equation}
which are sensitive to the existence of heavy particles in the Standard
Model, in particular the top quark and the Higgs boson. The dominant
contributions in the large top mass limit come from corrections to the $W$
and $Z$ boson propagators involving $t$ and $b$-quark loops. They are
proportional to $m_t^{2L}$, where $L$ is the number of loops, and
have recently been calculated up to three
loops~\cite{Faisst:2003px,vanderBij:2000cg}.

Another limit one can consider is the large Higgs mass limit. Here,
one might expect to find corrections of order $m_H^{2L}$, since the $W$
and $Z$ self-energies contain terms of this order, but it turns out
that such terms are screened in low energy observables such as the
$\rho$-parameter~\cite{Veltman:1976rt}. The leading
one-loop terms in $\Delta\rho$ depend logarithmically
on $m_H$~\cite{Longhitano:1980iz},
and the leading two-loop correction is proportional
to $m_H^2$~\cite{vanderBij:1983bw,Barbieri:1993ra}.
In this paper, we present the leading three-loop correction
$\Delta\rho^{(3)}$, which is of order $m_H^4$.

Radiative corrections to four-fermion processes include self-energy
corrections to the gauge boson propagators and vertex and box corrections.
In principle, all of these can affect $\rho$ and should be taken into
account. However, if the fermion
masses are neglected, the Higgs boson does not couple directly to the
fermions. In this case, the one-loop vertices and boxes
do not depend on $m_H$. Two and three-loop vertices and boxes
can only contribute to the leading order term
of $\Delta\rho^{(3)}$ if they contain one-loop subgraphs of order $m_H^2$
or two-loop subgraphs of order $m_H^4$. However, it is possible to choose
a renormalization scheme in which all terms of order $m_H^2$ ($m_H^4$)
in the relevant one-loop (two-loop) subgraphs
are absorbed into the renormalization
factors~\cite{vanderBij:1983bw,Einhorn:1988tc}. In such a scheme,
vertices and boxes do not contribute to
$\Delta\rho$ at the leading order in $m_H$, and one is left with just
the corrections coming from the transversal $W$ and $Z$ self-energies,
\begin{equation}
\Delta\rho = \frac{\Sigma_T^{ZZ}(0)}{M_Z^2} 
           - \frac{\Sigma_T^{WW}(0)}{M_W^2} \, .
\end{equation}

In section~\ref{sec:expansion}, we describe the calculation of 
the bare three-loop gauge boson self-energies, which we calculate
in the full electroweak Standard Model. The calculation is simplified
by the fact that they are only required at external momentum $p=0$,
which reduces the diagrams involved to vacuum diagrams. Diagrams
containing fermion loops are omitted, since after renormalization,
they do not give any contributions of order $m_H^4$.
We simplify the calculation further by systematically expanding all
diagrams in powers of $m_H$ for $m_H\gg M_W, M_Z$, using the method of
asymptotic expansions~\cite{smirnovbook}. In this way, all terms are factorized
into integrals depending only on the large scale $m_H$ on the one hand,
times integrals that depend on the small mass scales $M_W$ and $M_Z$
on the other. The factorized expressions are then reduced to a set of
independent master integrals using integration by parts identities. This
reduction is performed exactly in $d=4-\ep$ dimensions.
Higgs tadpole contributions and would-be Goldstone boson self-energies,
used to determine the renormalization constants, are calculated
by the same method. An attractive feature of this approach is
that it enables one to check Ward identities and observe cancellations
explicitly at the level of master integrals.

The renormalization is discussed in section~\ref{sec:renorm},
and the final result is presented in section~\ref{sec:result}.


\section{Large mass expansion and reduction to master integrals}
\label{sec:expansion}
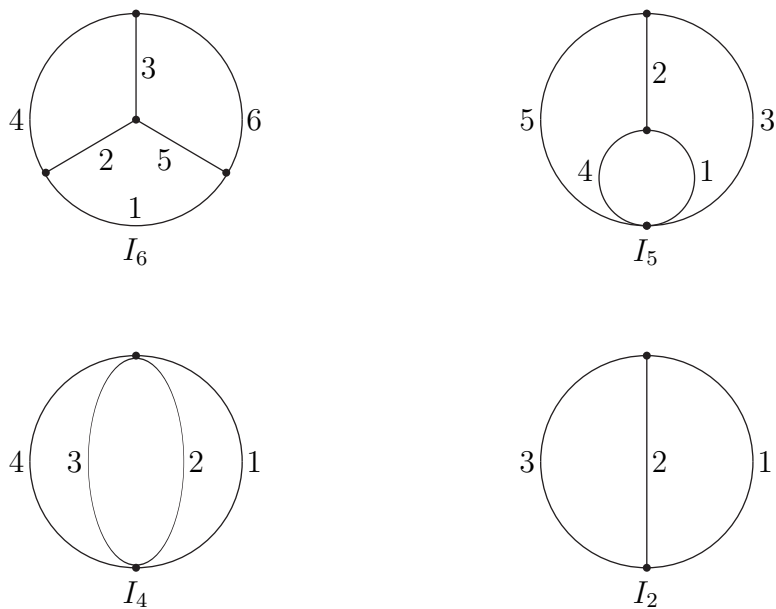
\begin{figure}[t]
    \begin{center}
    \begin{picture}(100,100)(0,0)
    \SetColor{Black}
    \CArc(50,50)(40,0,180)
    \SetColor{Black}
    \CArc(50,50)(40,180,360)
    \SetColor{Black}
    \Vertex(50,90){1.5} \Vertex(50,50){1.5}
    \Line(50,90)(50,50)
    \Vertex(16,30){1.5}
    \Line(16,30)(50,50)
    \Vertex(84,30){1.5}
    \Line(84,30)(50,50)
    \Text(55,70)[]{3}
    \Text(5,50)[]{4}
    \Text(95,50)[]{6}
    \Text(50,17)[]{1}
    \Text(39,35)[]{2}
    \Text(61,35)[]{5}
    \Text(50,0)[]{$I_{6}$}
    \end{picture}
    \hspace{3cm}
    \begin{picture}(100,100)(0,0)
    \SetColor{Black}
    \CArc(50,50)(40,0,180)
    \SetColor{Black}
    \CArc(50,50)(40,180,360)
    \SetColor{Black}
    \Vertex(50,90){1.5} \Vertex(50,46){1.5}
    \Line(50,90)(50,46)
    \Vertex(50,10){1.5}
    \CArc(50,28)(18,90,270)
    \Vertex(50,10){1.5}
    \CArc(50,28)(18,270,450)
    \Text(5,50)[]{5}
    \Text(96,50)[]{3}
    \Text(55,68)[]{2}
    \Text(27,31)[]{4}
    \Text(73,31)[]{1}
    \Text(50,0)[]{$I_{5}$}
    \end{picture}
\\[1cm]
    \begin{picture}(100,100)(0,0)
    \SetColor{Black}
    \CArc(50,50)(40,0,180)
    \SetColor{Black}
    \CArc(50,50)(40,180,360)
    \SetColor{Black}
    \Vertex(50,90){1.5} \Vertex(50,10){1.5}
    \Oval(50,50)(39,18)(0)
    \Text(27,50)[]{3}
    \Text(73,50)[]{2}
    \Text(95,50)[]{1}
    \Text(5,50)[]{4}
    \Text(50,0)[]{$I_{4}$}
    \end{picture}
    \hspace{3cm}
    \begin{picture}(100,100)(0,0)
    \SetColor{Black}
    \CArc(50,50)(40,0,180)
    \SetColor{Black}
    \CArc(50,50)(40,180,360)
    \SetColor{Black}
    \Vertex(50,90){1.5} \Vertex(50,10){1.5}
    \Line(50,90)(50,10)
    \Text(5,50)[]{3}
    \Text(55,50)[]{2}
    \Text(95,50)[]{1}
    \Text(50,0)[]{$I_{2}$}
    \end{picture}
    \end{center}
    \caption{The three-loop topologies $I_6$, $I_{5}$ and $I_{4}$,
    and the two-loop topology $I_{2}$}
    \label{figTopo}
\end{figure}

All our diagrams are generated by the program QGRAF~\cite{qgraf}.
The rest of the calculation is done mainly in FORM~\cite{FORM}.
For the three-loop one particle irreducible $W$ and $Z$ self-energies,
for example, there are 104340 and 82985 diagrams, respectively.
They can be divided into 80 different topologies. After setting
the external momentum $p=0$,
the diagrams can be expressed in terms of scalar vacuum integrals.
We use the following notations
\begin{eqnarray}
\label{eq:I6}
\lefteqn{
 I_6(m_1^2,m_2^2,m_3^2,m_4^2,m_5^2,m_6^2\,; n_1,n_2,n_3,n_4,n_5,n_6) =
}\hspace{1cm}  \nn \\ &&
        \int {d^dk_1\, d^dk_2\, d^dk_3} \,\,
              P{(k_{1}\,;\,m_{1})}^{n_1} 
              P{(k_2\,;\,m_{2})}^{n_2}
              P{(k_{3}\,;\,m_{3})}^{n_3} 
\nn \\ && \hspace{3mm}
              P{(k_{1}+k_{2}\,;\,m_{4})}^{n_4}
              P{(k_{2}+k_{3}\,;\,m_{5})}^{n_5} 
              P{(k_{1}+k_{2}+k_{3}\,;\,m_{6})}^{n_6}
\\
\label{eq:I5}
\lefteqn{
 I_5(m_1^2,m_2^2,m_3^2,m_4^2,m_5^2 \,; n_1,n_2,n_3,n_4,n_5) =
}\hspace{1cm}  \nn \\ &&
         \int {d^dk_1 \,d^dk_2\, d^dk_3}\,\,
               P{(k_{1}\,;\,m_{1})}^{n_1} 
               P{(k_2\,;\,m_{2})}^{n_2}
               P{(k_{3}\,;\,m_{3})}^{n_3} 
\nn \\ && \hspace{3mm}
               P{(k_{1}+k_{2}\,;\,m_{4})}^{n_4}
               P{(k_{2}+k_{3}\,;\,m_{5})}^{n_5}
\\   
\label{eq:I4Nf}
\lefteqn{
  I_{4}(m_1^2,m_2^2,m_3^2,m_4^2 \,; n_1,n_2,n_3,n_4) =
}\hspace{1cm}  \nn \\ &&
         \int {d^dk_1\, d^dk_2\, d^dk_3}\,\,
              {P{(k_{1}\,;\,m_{1})}}^{n_1} 
              {P{(k_2\,;\,m_{2})}}^{n_2}
              {P{(k_{3}\,;\,m_{3})}}^{n_3} 
\nn \\ && \hspace{3mm}
              {P{(k_{1}+k_{2}+k_{3}\,;\,m_{4})}}^{n_4}
\\ 
\label{eq:I2}
\lefteqn{
    I_2(m_1^2,m_2^2,m_3^2\,; n_1,n_2,n_3) =
}\hspace{1cm}  \nn \\ &&
      \int {d^dk_1\, d^dk_2 }\,\,
           P{(k_{1}\,;\,m_{1})}^{n_1} 
           P{(k_2\,;\,m_{2})}^{n_2}
           P{(k_{1}+k_{2}\,;\,m_{3})}^{n_3}
\\
\label{eq:I1}
\lefteqn{
       I_1(m_1^2\,; n_1) = 
}\hspace{1cm}  \nn \\ &&
         \int {d^dk_1}\,\,
         P{(k_{1}\,;\,m_{1})}^{n_1}                 
\end{eqnarray}
with \hspace{2mm} \[ P{(k\,;\,m)} = \frac{1}{k^2+m^2}\,. \] \\
These integrals correspond to the diagrams presented in Figure~\ref{figTopo}. 
In addition to the integrals~(\ref{eq:I6})--(\ref{eq:I1}), integrals with
scalar products $k_i\cdot k_j$ in the numerator appear.

In general, the propagators in the diagrams can depend on three different
non-zero masses: $m_H$, $M_W$ and $M_Z$. The Higgs mass also appears in
the scalar 3 and 4-point vertices via
$\lambda = g^2 m_H^2/(4 M_W^2)$.
We extract the leading $m_H$ dependent terms by performing an asymptotic
large mass expansion, considering $m_H$ to be large and $M_W$ and
$M_Z$ to be small. Here, we describe the procedure using the
language of expansion by regions~\cite{smirnovbook}.

The expansion is constructed by considering different regions in loop
momentum space, distinguished by the set of propagator momenta which are
large or small in those regions. In each region, a Taylor expansion of
all propagators in the small masses and in the small momenta of that
region is performed. Typically, the expansion generates
extra scalar products of loop momenta in the numerator
and higher powers of denominators, as compared to the original
diagrams. However, in each term, the dependence of the integrand on
the large and small parameters is factorized. The resulting expression
is then integrated over the whole loop momentum space.

For the three-loop vacuum topology $I_6$, there are 15
regions in loop momentum space to consider.\\[1em]
1. The region where the momenta in all propagators are large. In
this case, the Taylor expansion yields three-loop vacuum integrals
in which all masses are either zero, or equal to $m_H$.\\[1em]
2. Six regions where one momentum, e.g. $k_1$, is small, while the others
are large. Here, the Taylor expansion leads to products of two-loop
vacuum integrals depending on $m_H$, times one-loop vacuum integrals
depending on $M_W$ or $M_Z$.\\[1em]
3. Three regions where the momenta in two non-adjacent propagators,
e.g. $k_1$ and $k_3$, are small, and the momenta in the other
propagators are large.\\[1em]
4. Four regions where the momenta in three propagators that are connected
to a common vertex, e.g. $k_1$, $k_2$ and $k_1+k_2$, are small, and the
momenta in the other propagators are large. In these regions and in the
regions of type 3, the Taylor expansion leads to products of one-loop
vacuum integrals depending on $m_H$, times two-loop vacuum integrals
depending on $M_W$ and $M_Z$.\\[1em]
5. The region where all momenta are small. In this region, the
Taylor expansion gives three-loop vacuum integrals
depending only on $M_W$ and $M_Z$. These two-scale integrals
are the most complicated ones that appear in the expansion. So
far, only a few of the corresponding master integrals have been
calculated in the literature~\cite{Davydychev:2003mv}.
\\[1em]

Not all regions give non-vanishing contributions for all diagrams. In many
cases, after the Taylor expansion we are left with scale-less integrals,
which are zero in dimensional regularization. Which regions actually do
contribute depends on the distribution of large masses in the diagrams
concerned. For example, in case 2 above, the region where $k_1$ is small
only gives a non-vanishing contribution when $m_1$ is a small mass.

Obviously, the factorizable integrals coming from regions 2, 3 and 4
are easy to deal with, since they are simply products of one and
two-loop vacuum integrals.
The $I_6$ integrals
coming from region 1 can be classified into ten different kinds, according
to the distribution of massless and massive denominators they contain. For
two of these categories, we follow the integration-by-parts~\cite{ibp}
reduction algorithm obtained by Broadhurst~\cite{Broadhurst:1991fi},
to reduce them to master integrals. Reductions for the eight
other categories have been indicated by Avdeev~\cite{Avdeev:1995eu},
but here, we prefer to use our own reduction routines.
As a cross-check, we have also
performed the reduction of these single-scale integrals using
the automatic integral reduction package AIR~\cite{Anastasiou:2004vj}.
The master integrals themselves are known, some in terms of $\Gamma$
functions, the others as expansions in
$\ep$~\cite{Broadhurst:1998rz,Fleischer:1999mp}.

The three-loop integrals from region 5 appear to present a more difficult
problem, since some of them depend on {\em two} mass scales: $M_W$
and $M_Z$. It turns out that, in the three-loop gauge boson
self-energies up to the order $m_H^4$, they are all of the
$I_{4}$ topology (sometimes with scalar products in the numerator).
However, by using symmetry relations between such integrals, we find
that they cancel out of the gauge boson self-energies at this order
in $m_H$, provided we sum over all diagrams.

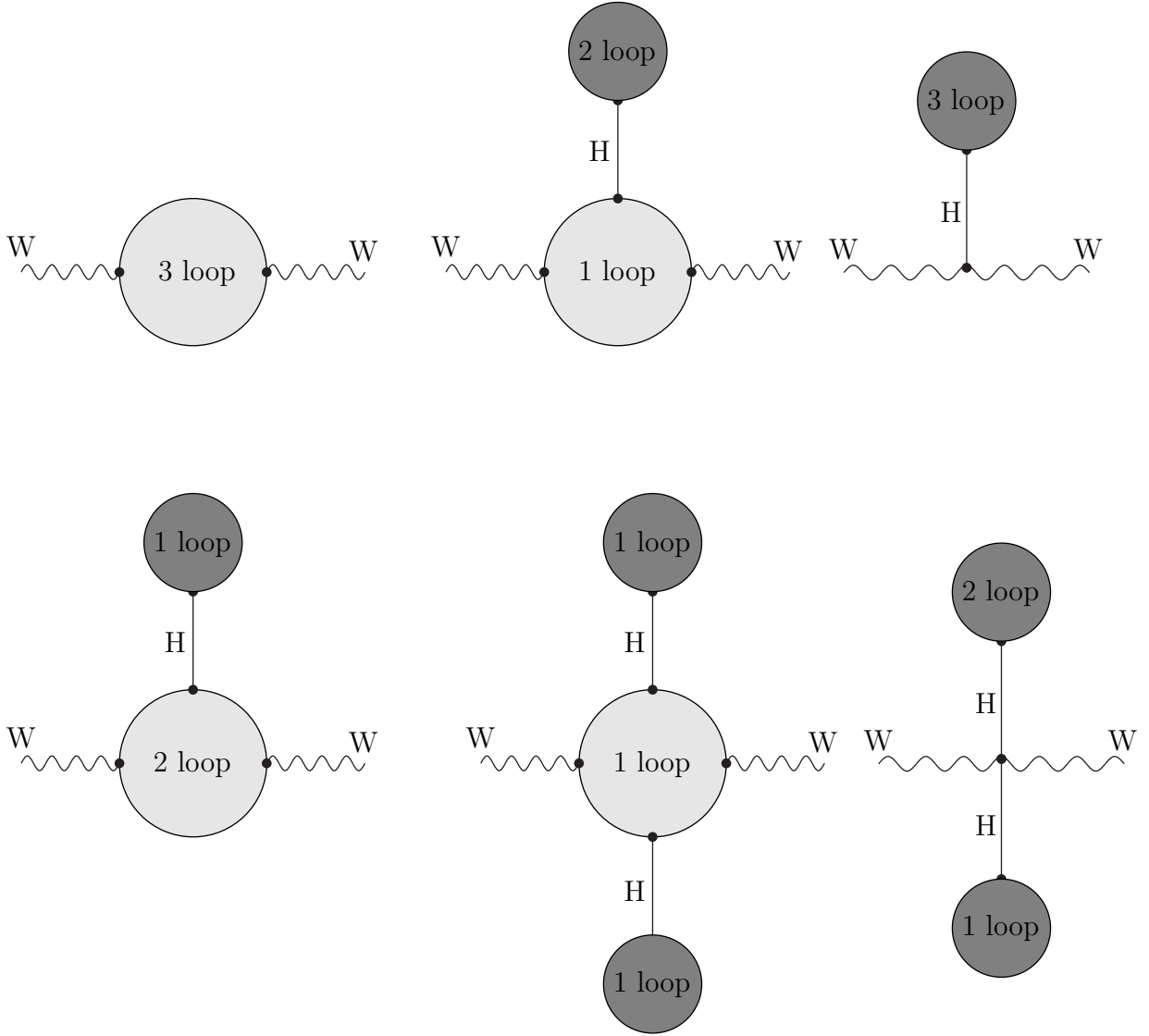
\begin{figure}
   \begin{picture}(140,200)(30,0)
    \SetColor{Black}
    \GCirc(100,100){30}{0.9}
    \SetColor{Black}
    \Photon(30,100)(70,100){3}{4}  \Vertex(70,100){2}
    \Photon(130,100)(170,100){3}{4} \Vertex(130,100){2}
    \Text(30,111)[]{W}
    \Text(170,109)[]{W}
    \Text(100,100)[]{ 3 loop}
    \end{picture}
\hspace{1cm}
   \begin{picture}(140,200)(30,0)
    \SetColor{Black}
    \GCirc(100,100){30}{0.9} 
    \SetColor{Black}
    \SetColor{Black}
    \Photon(30,100)(70,100){3}{4}  \Vertex(70,100){2}
    \Photon(130,100)(170,100){3}{4} \Vertex(130,100){2}
    \Vertex(100,130){2}
    \Line(100,130)(100,170)
    \Vertex(100,170){2}
    \GCirc(100,190){20}{0.5}
    \Text(30,111)[]{W}
    \Text(170,109)[]{W}
    \Text(93,150)[]{H}
    \Text(100,190)[]{ 2 loop }
    \Text(100,100)[]{ 1 loop }
    \end{picture}  
\nolinebreak
   \begin{picture}(140,200)(30,-50)
    \SetColor{Black}
    \Photon(50,50)(150,50){3}{6.5}
    \Vertex(100,52){2}
    \SetColor{Black}
    \Line(100,50)(100,100)
    \Vertex(100,100){2}
    \SetColor{Black}
    \GCirc(100,120){20}{0.5}
    \Text(94,75)[]{H}
    \Text(50,60)[]{W}
    \Text(150,60)[]{W}
    \Text(100,120)[]{3 loop}
   \end{picture}  
\\
    \begin{picture}(140,200)(30,0)
    \SetColor{Black}
    \GCirc(100,100){30}{0.9} 
    \SetColor{Black}
    \SetColor{Black}
    \Photon(30,100)(70,100){3}{4}  \Vertex(70,100){2}
    \Photon(130,100)(170,100){3}{4} \Vertex(130,100){2}
    \Vertex(100,130){2}
    \Line(100,130)(100,170)
    \Vertex(100,170){2}
    \GCirc(100,190){20}{0.5}
    \Text(30,111)[]{W}
    \Text(170,109)[]{W}
    \Text(93,150)[]{H}
    \Text(100,190)[]{ 1 loop }
    \Text(100,100)[]{ 2 loop }
    \end{picture}  
\hspace{1.5cm}
   \begin{picture}(140,200)(30,0)
    \SetColor{Black}
    \GCirc(100,100){30}{0.9} 
    \SetColor{Black}
    \Photon(30,100)(70,100){3}{4}  \Vertex(70,100){2}
    \Photon(130,100)(170,100){3}{4} \Vertex(130,100){2}
    \Vertex(100,130){2}
    \Line(100,130)(100,170)
    \Vertex(100,170){2}
    \GCirc(100,190){20}{0.5}
    \Vertex(100,70){2}
    \Line(100,70)(100,30)
    \GCirc(100,10){20}{0.5} 
    \Text(30,111)[]{W}
    \Text(170,109)[]{W}
    \Text(93,150)[]{H}
    \Text(100,190)[]{ 1 loop }
    \Text(100,10)[]{ 1 loop }
    \Text(93,48)[]{H}
    \Text(100,100)[]{ 1 loop }
    \end{picture}  
\nolinebreak
   \begin{picture}(140,200)(30,-50)
    \Photon(50,50)(150,50){3}{6.5}
    \Vertex(100,52){2}
    \SetColor{Black}
    \Line(100,50)(100,100)
    \Vertex(100,100){2}
    \SetColor{Black}
    \GCirc(100,120){20}{0.5}
    \Text(100,120)[]{2 loop}
    \Text(94,75)[]{H}
    \Text(50,60)[]{W}
    \Text(150,60)[]{W}
    \Vertex(100,52){2}
    \SetColor{Black}
    \Line(100,50)(100,0)
    \Vertex(100,3){2}
    \SetColor{Black}
    \GCirc(100,-17){20}{0.5}
    \Text(100,-17)[]{1 loop}
    \Text(94,25)[]{H}
    \end{picture}  
    \caption{ The full $WW$ three-loop self-energy,
             the light-grey blobs represent the sum of one particle 
             irreducible one, two and three-loop diagrams,
             the dark-grey blobs represent the full one, 
             two and three loop tadpoles. }
    \label{figWW3LSE}
\end{figure}

\begin{figure}
   \begin{picture}(140,200)(30,18)
    \SetColor{Black}
    \Line(100,50)(100,100)
    \Vertex(100,100){2}
    \SetColor{Black}
    \GCirc(100,120){20}{0.9}
    \Vertex(100,140){2}
    \Line(100,140)(100,170)
    \GCirc(100,190){20}{0.9}
    \Text(100,120)[]{1 loop}
    \Text(100,190)[]{1 loop}
    \Text(93,75)[]{H}
    \Text(93,155)[]{H}
    \end{picture}  
\begin{picture}(140,200)(30,18)
    \SetColor{Black}
    \Line(100,50)(100,100)
    \Vertex(100,100){2}
    \SetColor{Black}
    \Line(100,100)(136,136) 
    \Vertex(136,136){2}   
    \Line(100,100)(64,136)
    \Vertex(64,136){2} 
    \GCirc(150,150){20}{0.9}
    \GCirc(50,150){20}{0.9}
    \Text(93,75)[]{H}
    \Text(150,150)[]{1 loop}
    \Text(50,150)[]{1 loop}
    \end{picture}
   \begin{picture}(140,200)(30,18)
    \SetColor{Black}
    \SetColor{Black}
    \Line(100,50)(100,100)
    \Vertex(100,100){2}
    \SetColor{Black}
    \GCirc(100,120){20}{0.9}
    \Text(94,75)[]{H}
    \Text(100,120)[]{2 loop}   
    \end{picture}
    \caption{ The full two loop Higgs tadpole. }
    \label{fig2Ltad}  
\end{figure}
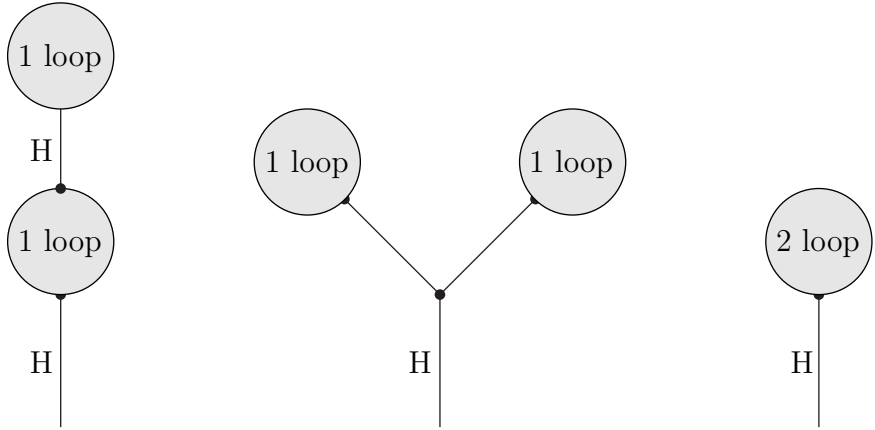

The gauge boson self-energies we need are the ``full'' self-energies,
consisting of a one-particle irreducible part plus terms containing Higgs
tadpole insertions, which, in turn, include tadpole insertions themselves,
as shown in Figures~\ref{figWW3LSE} and~\ref{fig2Ltad}.

We have checked several Ward identities that are satisfied by the
unrenormalized self-energies. In particular, we have checked
that the photon self-energy at zero momentum
$\Sigma_T^{AA,3-loop,full}(0)$ vanishes at order $m_H^4$.
For the would-be Goldstone bosons we have checked that
$\Sigma^{\phi\phi,3-loop,full}(0)$ and
$\Sigma^{\phi^0\phi^0,3-loop,full}(0)$
both vanish at order $m_H^6$. These identities are only
valid for the self-energies including tadpole contributions.

\section{Renormalization}
\label{sec:renorm}
We write the Standard Model Lagrangian without fermions as
${\cal L}={\cal L}_{inv}+{\cal L}_{fix}+{\cal L}_{FP}$.
The invariant part is
\begin{equation}
\label{eq:Linv}
{\cal L}_{inv} = -\frac{1}{4} W^a_{\mu\nu} W^{a,\mu\nu}
           -\frac{1}{4} B_{\mu\nu} B^{\mu\nu}
           -\left( D_{\mu} \Phi \right)^{\dagger}
            \left( D^{\mu} \Phi \right)
           -\frac{1}{2} \lambda {\left(\Phi^{\dagger}\Phi\right)}^2 
           -\mu \, \Phi^{\dagger}\Phi
\end{equation}
where $W^a_{\mu\nu}$ and $B^{\mu\nu}$ denote the curvatures 
of the $SU(2)_L$ and $U(1)_Y$ gauge fields $W^a_{\mu}$ and $B^{\mu}$,
\begin{equation}
\Phi= \frac{1}{\sqrt{2}} \left(
 \begin{array}{c}
 H+\sqrt{2}v + i \phi^0
\\
 i \phi^1 - \phi^2
 \end{array}
\right)
\end{equation}
is the Higgs doublet, and
\begin{equation}
D_{\mu} \Phi = \left( \partial_{\mu}
 - \frac{i\,g}{2} W^a_{\mu} \tau^a
 - \frac{i\,g'}{2} B_{\mu}
\right) \Phi 
\end{equation}
its covariant derivative. The fields $\phi^{\pm}$,
$W_{\mu}^{\pm}$, $Z_{\mu}$ and $A_{\mu}$ are defined by
\begin{eqnarray}
\phi^{\pm} & = & \frac{1}{\sqrt{2}} \( \phi^1 \mp i \phi^2 \) \, ,
\\
W_{\mu}^{\pm} & = & \frac{1}{\sqrt{2}} \( W_{\mu}^1 \mp i W_{\mu}^2 \) \, ,
\\
Z_{\mu} & = & c_W W_{\mu}^{3} - s_W B_{\mu} \, ,
\\
A_{\mu} & = & s_W W_{\mu}^{3} + c_W B_{\mu} \, ,
\end{eqnarray}
and the coupling constants $g'$ and $g$ are related to each other by
\begin{eqnarray}
g' & = & - g \frac{s_W}{c_W} \, .
\end{eqnarray}

The gauge fixing term is given by
\begin{equation}
\label{eq:Lfix}
{\cal L}_{fix} = -C^+ C^- - \frac{1}{2} {(C^Z)}^2 - \frac{1}{2} {(C^A)}^2
\end{equation}
with
\begin{eqnarray}
\label{eq:Cpm}
C^{\pm} & = & - \partial_{\mu} W^{\pm,\mu} + \xi M_W \phi^{\pm}
\\
\label{eq:Cz}
C^{Z} & = & - \partial_{\mu} Z^{\mu} + \xi M_Z \phi^{0}
\\
C^{A} & = & - \partial_{\mu} A^{\mu}
\end{eqnarray}
In equations~(\ref{eq:Cpm})--(\ref{eq:Cz}), we have introduced
a gauge parameter $\xi$. The standard 't~Hooft Feynman gauge
fixing term corresponds to $\xi=1$.
Finally, the Faddeev Popov
Lagrangian ${\cal L}_{FP}$ is derived
from the variation of ${\cal L}_{fix}$ under gauge transformations.

Because the tree level $\rho$-parameter is independent of
the parameters in the Lagrangian,
$\Delta\rho^{(3)}$ is not affected by any three-loop counterterms.
Therefore, we only need
to renormalize the model up to two loops. Our renormalization scheme is
similar to the one used in ref.~\cite{vanderBij:1983bw}. However, in line
with our calculation of the unrenormalized diagrams, the renormalization
factors are systematically expanded in powers of $m_H$, and only terms
that can give a contribution of order $m_H^4$ to $\Delta\rho^{(3)}$
are retained, i.e. terms of order $m_H^2$ in one-loop counterterms,
and terms of order $m_H^4$ in two-loop counterterms. This implies that
we do not renormalize $g$ or $c_W$ at all.

In the Lagrangian, we express $\lambda$, $\mu$, $v$ and $M_Z$
in terms of the four independent parameters
$g$, $c_W$, $m_H$ and $M_W$
by
$\mu = -\frac{1}{2} m_H^2$,
$\lambda = g^2 m_H^2/(4 M_W^2)$,
$v = \sqrt{2} \frac{M_W}{g}$,
and $M_Z$ = $M_W/c_W$.
Then, the masses $m_H$ and $M_W$, and the Higgs and would-be Goldstone fields
are replaced with bare masses and fields
\begin{eqnarray}
\label{eq:mren}
m_H^{(0)} & = & Z_m \, m_H
\\
\label{eq:Mren}
M_W^{(0)} & = & Z_M \, M_W
\\
\label{eq:Hren}
H^{(0)} & = & Z_H \, H
\\
\label{eq:phiren}
\phi^{\pm,(0)} & = & Z_H \phi^{\pm}
\\
\label{eq:phi0ren}
\phi^{0,(0)} & = & Z_H \phi^{0}
\end{eqnarray}
The gauge parameter $\xi$ is renormalized in such a way
as to compensate the effect of the above renormalizations in
the gauge fixing term:
\begin{equation}
\xi^{(0)} = Z_H^{-1} Z_M^{-1} \, \xi
\end{equation}
We choose the renormalized $\xi$ to be equal to one.
Each $Z$-factor is written as
\begin{equation}
Z = 1 - \delta^{(1)} - \delta^{(2)} \, .
\end{equation}

The renormalization constants $Z_M$ and $Z_H$ are fixed by imposing
conditions on the renormalized $W$ and $\phi$ self-energies
\begin{equation}
\label{eq:ctMcondition}
\Sigma_T^{WW,ren}|_{p^2=0} \, \sim \, 0
\end{equation}
and
\begin{equation}
\label{eq:ctHcondition}
\frac{\partial}{\partial p^2}
\Sigma^{\phi\phi,ren}|_{p^2=0} \, \sim \, 0 \, ,
\end{equation}
where the notation $X \sim 0$ means that $X$ does not
contain any terms of order $m_H^2$ at one loop, or of
order $m_H^4$ at two loops.
These renormalizations remove all the $m_H^2$ and $m_H^4$ terms from the
one and two-loop gauge boson self-energies, the $\phi$ self-energies,
and the mixings between $\phi$'s and gauge bosons.
The Higgs mass renormalization constant $Z_m$ is fixed by
demanding that
\begin{equation}
\label{eq:ctmcondition}
\frac{ \mbox{Re} \, \Sigma^{HH,ren}|_{p^2+m_H^2=0}} {m_H^2} \, \sim \, 0 \, .
\end{equation}

Solving equations~(\ref{eq:ctMcondition})--(\ref{eq:ctHcondition}), we find
the following expressions in terms of vacuum integrals.
\begin{eqnarray}
\label{eq:ctH1}
\delta_H^{(1)} & = & \frac{1}{i{(2\pi)}^d} \frac{g^2}{M_W^2}
                     I_1(m_H^2;1) \, \frac{\ep}{8\,(\ep-4)}
\\
\label{eq:ctM1}
\delta_M^{(1)} & = & \frac{1}{i{(2\pi)}^d} \frac{g^2}{M_W^2}
                     I_1(m_H^2;1) \, \frac{6-\ep}{4\,(\ep-4)}
\\
\label{eq:ctH2}
\delta_H^{(2)} & = & {\left[
                     \frac{1}{i{(2\pi)}^d} \frac{g^2}{M_W^2}
                     \right]}^2
\left\{
I_1(m_H^2;1)^2
          \left( 
            \frac{13}{128}
          + \frac{3\,\ep}{64}
          + \frac{3}{16\,(\ep-4)}
          + \frac{1}{8\,(\ep-4)^2}
          \right)
\right. \nonumber \\ &&
       {} + m_H^2 \, I_2(m_H^2,m_H^2,m_H^2;1,1,1)
          \left(
            \frac{3\,\ep}{64}
          + \frac{9}{64}
          + \frac{3}{8\,(\ep-4)}
          \right)
\nonumber \\ && \left.
       {} + m_H^2 \, I_2(m_H^2,0,0;1,1,1)
          \left(
            \frac{\ep}{64}
          + \frac{11}{64}
          + \frac{3}{4\,(\ep-4)}
          \right)
\right\}
\nonumber \\ &&
       {} + \frac{1}{i{(2\pi)}^d} \frac{g^2}{M_W^2}
          I_1(m_H^2;1)
          \left(
            \frac{1}{4}
          + \frac{\ep}{8}
          + \frac{1}{(\ep-4)}
          \right) \delta_m^{(1)}
\\
\label{eq:ctM2}
\delta_M^{(2)} & = & {\left[
                     \frac{1}{i{(2\pi)}^d} \frac{g^2}{M_W^2}
                     \right]}^2
\left\{
I_1(m_H^2;1)^2
          \left(
            \frac{13}{32}
          - \frac{3\,\ep}{32}
          + \frac{1}{8\,(\ep-4)^2}
          \right)
\right. \nonumber \\ &&
       {} + m_H^2 \, I_2(m_H^2,m_H^2,m_H^2;1,1,1)
          \left(
          - \frac{3\,\ep}{32}
          + \frac{3}{8}
          + \frac{3}{8\,(\ep-4)}
          \right)
\nonumber \\ && \left.
       {} + m_H^2 \, I_2(m_H^2,0,0;1,1,1)
          \left(
          - \frac{\ep}{32}
          + \frac{5}{16}
          + \frac{3}{4\,(\ep-4)}
          \right)
\right\}
\nonumber \\ &&
       {} + \frac{1}{i{(2\pi)}^d} \frac{g^2}{M_W^2}
          I_1(m_H^2;1)
          \left(
            1
          - \frac{\ep}{4}
          + \frac{1}{(\ep-4)}
          \right) \delta_m^{(1)}
\end{eqnarray}

In order to determine the renormalization constant $\delta_m^{(2)}$,
we use analytical results for the two-loop on-shell Higgs self-energy from
ref.~\cite{Borodulin:1996br}. Some care is needed here, because the
bare parameters $M_{W0}$ and $m_{H0}$ used in ref.~\cite{Borodulin:1996br}
do not correspond exactly to our bare parameters $m_H^{(0)}$, $M_W^{(0)}$.
This is due to the fact that we do not introduce a counterterm $\delta v^2$
for the Higgs tadpole, but instead, explicitly include tadpole contributions.
Their $m_{H0}/m_H$ corresponds to our
$Z_m Z_H / Z_M$. Taking this difference into account, we find
\begin{eqnarray}
\label{eq:ctm1}
\delta_m^{(1)} & = &
                   {\left(\frac{g^2}{16\pi^2}\right)}
                   {\left(\frac{m_H^2}{4\pi}\right)}^{-\frac{\ep}{2}}
                   \Gamma\left(1+\frac{\ep}{2}\right)
                   \frac{m_H^2}{M_W^2}
\left\{
          - \frac{3}{4\,\ep}
          - \frac{9}{8}
          + \frac{3}{16}\pi\sqrt{3}
\right. \nonumber \\ && \left.
      + \ep \left(
          - \frac{3}{32}\pi\sqrt{3}\log{3}
          + \frac{3}{16}\pi\sqrt{3}
          + \frac{1}{16}\pi^2
          + \frac{3}{8}\sqrt{3}\,C
          - \frac{21}{16}
           \right)
\right\} \, ,
\\
\label{eq:ctm2}
\delta_m^{(2)} & = &
                   {\left(\frac{g^2}{16\pi^2}\right)}^2
                   {\left(\frac{m_H^2}{4\pi}\right)}^{-\ep}
                   \Gamma^2\left(1+\frac{\ep}{2}\right)
                   \frac{m_H^4}{M_W^4}
\left\{
          - \frac{45}{32\,\ep^2}
          +\frac{1}{\ep} \left(
            \frac{27}{64}\pi\sqrt{3}
          - \frac{189}{64}
            \right)
\right. \nonumber \\ && \left.
          - \frac{807}{256}
          - \frac{27}{128}\pi\sqrt{3}\log{3}
          + \frac{57}{32}\pi\sqrt{3}
          - \frac{39}{32}\pi\,C
          - \frac{261}{512}\pi^2
          - \frac{9}{32}\sqrt{3}\,C
          + \frac{63}{32}\zeta(3)
\right\} \, .
\nonumber \\
\end{eqnarray}

The correction to the $\rho$-parameter can now be written as
\begin{equation}
\label{eq:rho3sigma}
\Delta\rho^{(3)} = \frac{\Sigma_T^{ZZ,3-loop,ren}(0)}{M_Z^2} 
                 - \frac{\Sigma_T^{WW,3-loop,ren}(0)}{M_W^2} \, .
\end{equation}
The counterterms $\delta^{(1)}$, $\delta^{(2)}$ are substituted
into equation (\ref{eq:rho3sigma}) in two steps. In the first
step,
$\delta_H^{(1)}$, 
$\delta_H^{(2)}$, 
$\delta_M^{(1)}$ and
$\delta_M^{(2)}$
are replaced with the expressions (\ref{eq:ctH1})--(\ref{eq:ctM2}),
(keeping $\delta_m^{(1)}$ and $\delta_m^{(2)}$ as symbols).
This step is done before performing any expansions in $\ep$.
At this point, all master integrals depending on $M_W$ or $M_Z$,
which originate from small momenta regions, cancel out exactly,
so that only master integrals depending on $m_H$ are left.
In the second step, the counterterms $\delta_m^{(1)}$ and
$\delta_m^{(2)}$ are replaced with
(\ref{eq:ctm1})--(\ref{eq:ctm2})
and the master integrals are expanded in $\ep$, using
results from refs.~\cite{Broadhurst:1998rz,Fleischer:1999mp}.
In the expansions, singular terms of order $1/\ep^j$, $j=1,2,3,4$,
appear and cancel each other. 
The order $\ep$ term of $\delta_m^{(2)}$ is not needed,
because the coefficient of $\delta_m^{(2)}$ in $\Delta\rho^{(3)}$
is finite in the limit $\ep \to 0$. Similarly,
$\delta_m^{(1)}$ is not needed beyond the term of
order $\ep$ given in eq.~(\ref{eq:ctm1}).

As a check on the renormalization procedure, we have verified that
the renormalized three-loop photon self-energy,
the photon-$Z$ mixing self-energy,
and the $\phi$ self-energies vanish at zero external momentum.


\section{Results and conclusion}
\label{sec:result}
Finally, combined with the one-loop~\cite{Longhitano:1980iz}
and two-loop~\cite{vanderBij:1983bw} terms, the three-loop
correction to the $\rho$-parameter reads
\begin{equation}
\label{eq:rho123}
\rho = 1 + \Delta\rho^{(1)} + \Delta\rho^{(2)} + \Delta\rho^{(3)} \, ,
\end{equation}
with
\begin{eqnarray}
\label{eq:rho}
\Delta\rho^{(1)}    & = & - \frac{3}{4} \frac{g^2}{16\pi^2}
                   \frac{s_W^2}{c_W^2}
                   \log\left(\frac{m_H^2}{M_W^2}\right) \, ,
\\
\Delta\rho^{(2)} & = & {\left(\frac{g^2}{16\pi^2}\right)}^2
                   \frac{s_W^2}{c_W^2}
                   \frac{m_H^2}{M_W^2}
                   \left(\, -\frac{21}{64} + \frac{9}{32}\pi\sqrt{3}
                        + \frac{3}{32}\pi^2 - \frac{9}{8} C \sqrt{3}
                  \, \right) 
\nn\\
                & = & {{\left(\frac{g^2}{16\pi^2}\right)}^2
                   \frac{s_W^2}{c_W^2}
                   \frac{m_H^2}{M_W^2}}\, \(\, 0.1499\,\) \, ,
\\
\Delta\rho^{(3)} & = & 
               {{\left(\frac{g^2}{16\pi^2}\right)}^3
                   \frac{s_W^2}{c_W^2}
                   \frac{m_H^4}{M_W^4}} 
         \(\, - \frac{21}{512}
          + \frac{729}{512}\,\pi \sqrt{3}
          - \frac{3391}{4608}\, \pi^2
          - \frac{9}{16}\pi\, C 
\right.\nn \\ && \hspace{35mm}
          - \frac{1577}{2304} \,\pi^3 \sqrt{3}
          - \frac{9109}{69120}\, \pi^4
          + \frac{99}{16}\, \sqrt{3} \, \log{3} \,\,C
\nn \\ && \hspace{35mm}
          - \frac{297}{32} \sqrt{3}\, {\mbox{Ls}}_3{(2 \pi/3)}
          - \frac{399}{16} \sqrt{3}\, C
          + \frac{3043}{128}\, \zeta{(3)}
\nn \\ && \hspace{35mm} \left.
          + \frac{567}{32}\, C^2
         + \frac{109}{8}\,U_{3,1}
          - 36 \, V_{3,1}\, \) 
\nn \\  &=& {{\left(\frac{g^2}{16\pi^2}\right)}^3
                   \frac{s_W^2}{c_W^2}
                   \frac{m_H^4}{M_W^4}}\, \(\,-1.7282\,\)  \, .
\end{eqnarray}
The constants appearing in $\Delta\rho^{(3)}$ are defined
by~\cite{Broadhurst:1998rz,Fleischer:1999mp}
\begin{eqnarray}
U_{3,1} &=& \frac{1}{2} \zeta(4) + \frac{1}{2} \zeta(2) \log^2 2
           - \frac{1}{12} \log^4 2 - \Li{4}{\frac{1}{2}} 
\nn\\
        &=& -0.11787599965
\\
V_{3,1} &=& 
\sum_{m>n>0}\frac{(-1)^m\cos(2\pi n/3)}{m^3n}
\nn\\
       &=& -0.03901272636
\\ C &=& {\mbox{Cl}}_2\left(\pi/3\right)
\end{eqnarray}
The log-sine integral is defined by
\begin{equation}
\Ls{3}{\theta} = - \int_0^{\theta} \mbox{d} \phi \;
\log^{2} \left| 2 \sin\frac{\phi}{2} \right| .
\end{equation}

\begin{figure}
\begin{center}
\includegraphics[]{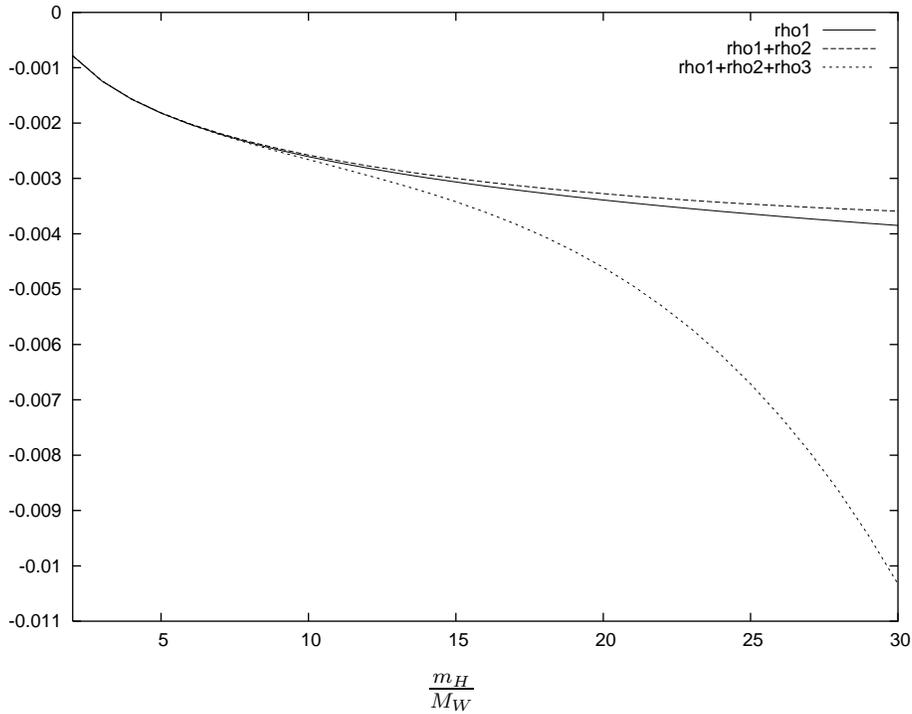}\\
$\frac{m_H}{M_W}$
\end{center}
\caption{
The combined effect of $\Delta\rho^{(1)}$, $\Delta\rho^{(2)}$ and
$\Delta\rho^{(3)}$ as a function of $m_H/M_W$.}
\label{fig:rhoplot}
\end{figure}

Some numerical values are shown in Table~1 and in
Figure~\ref{fig:rhoplot}, where we have
used
\begin{equation}
g^2 = \frac{e^2}{s_W^2} = \frac{4\pi\alpha}{s_W^2}
\end{equation}
for the weak coupling constant, with
$\alpha=1/137$
and
$s_W^2 = 0.23$.
While $\Delta\rho^{(3)}$ is very small for low values of $m_H$,
it soon becomes more important than $\Delta\rho^{(2)}$.
The two contributions cancel each other
at $m_H\approx 6 M_W\approx 480~\mbox{GeV}$. $\Delta\rho^{(3)}$
becomes equal to $\Delta\rho^{(1)}$ for $m_H\approx 2~\mbox{TeV}$.

Comparing our result for the large Higgs mass limit with
the ones of the large top quark mass limit~\cite{Faisst:2003px},
we can say that for $m_H\approx 500~\mbox{GeV}$, $\Delta\rho^{(3)}$
is already larger than the three-loop pure electroweak correction of
order $M_t^6$, which only grows as $m_H^2 \log(m_H^2/M_t^2)$
in the large $m_H$ limit. However, it is still very small
compared with the mixed electroweak/QCD correction term of order
$\alpha_s M_t^4$.

\begin{table}
\[\begin{array}{|c|c|c|c|}  \hline 
     m_H/M_W&\Delta\rho^{(1)}&\Delta\rho^{(2)}&\Delta\rho^{(3)}
     \vphantom{\overline{\overline{\Delta\rho^{(3)}}}}
\\[2pt]\hline
2&{\tt-0.00078}&{\tt1.14\,10^{-6}}&
{\tt-1.33\,10^{-7} }
\\[3pt]\hline
5&{\tt-0.0018}&{\tt7.14\,10^{-6}}&
{\tt-5.20\,10^{-6} }
\\[3pt]\hline
6&{\tt-0.0020}&{\tt0.000010}&
{\tt-0.000011 }
\\[3pt]\hline
7&{\tt-0.0022}&{\tt0.000014}&
{\tt-0.000020 }
\\[3pt]\hline
8&{\tt-0.0024}&{\tt0.000018}&
{\tt-0.000034 }
\\[3pt]\hline
9&{\tt-0.0025}&{\tt0.000023}&
{\tt-0.000055 }
\\[3pt]\hline
10&{\tt-0.0026}&{\tt0.000029}&
{\tt-0.000083}
\\[3pt]\hline
15&{\tt-0.0031}&{\tt0.000064}&
{\tt-0.00042}
\\[3pt]\hline
20&{\tt-0.0034}&{\tt0.00011}&
{\tt-0.0013}
\\[3pt]\hline
25&{\tt-0.0036}&{\tt0.00018}&
{\tt-0.0032}
\\[3pt]\hline
26&{\tt-0.0037}&{\tt0.00019}&
{\tt-0.0038}
\\[3pt]\hline
27&{\tt-0.0037}&{\tt0.00021}&
{\tt-0.0044}
\\[3pt]\hline
28&{\tt-0.0038}&{\tt0.00022}&
{\tt-0.0051}
\\[3pt]\hline
29&{\tt-0.0038}&{\tt0.00024}&
{\tt-0.0059}
\\[3pt]\hline
30&{\tt-0.0038}&{\tt0.00026}&
{\tt-0.0067}
\\[3pt]\hline
\end{array}\]
\caption{Corrections to ${\rho}$ as a function 
         of \,\,${{m_H}/{M_W}}$}
\end{table}

The original question that motivated this calculation was, whether
inclusion of the three-loop corrections with strong interactions could
lead to an effect mimicking the one-loop effects of a light Higgs boson.
The result of this calculation shows that this is highly unlikely.
The sign of the three-loop correction is the same as the one-loop
correction. Therefore with increasing Higgs-mass the three-loop term only
makes the effect grow faster, instead of the three-loop term partially
cancelling the one-loop correction. Therefore the presence of a strongly
interacting heavy Higgs-sector appears to be extremely unlikely, and
the data can indeed be used as a strong indication for a light Higgs
boson sector.


\subsection*{Acknowledgements}

We are grateful to G.~Jikia for helpful discussions and to P.~Marquard for
his assistance with some cross-checks. This work was supported by the
DFG-Forschergruppe "Quantenfeldtheorie, Computeralgebra and Monte-Carlo
Simulation". It was also supported by the European Union
under contract HPRN-CT-2000-00149.


\end{document}